\documentclass[prb,aps]{revtex4}
\usepackage{epsfig}

\newcommand{\vk}{{\mbox{\boldmath$k$}}}

\newcommand{\veta}{{\mbox{\boldmath$\eta$}}}

\newcommand{\vupsilon}{\mbox{\boldmath$\upsilon$}}

\newcommand{\vf}{\mbox{\boldmath$f$}}

\newcommand{\vq}{\mbox{\boldmath$q$}}

\newcommand{\vsig}{\mbox{\boldmath$\sigma$}}
\newcommand{\vd}{\mbox{\boldmath$d$}}
\newcommand{\mhx}{\hat{\mbox{\boldmath$x$}}}
\newcommand{\mhy}{\hat{\mbox{\boldmath$y$}}}
\newcommand{\mhz}{\hat{\mbox{\boldmath$z$}}}

\newcommand{\vP}{\mbox{\boldmath$P$}}
\newcommand{\vQ}{\mbox{\boldmath$Q$}}

\begin{document}
\title{Microscopic theories for cubic
and tetrahedral superconductors: application to PrOs$_4$Sb$_{12}$
}
\author{S. Mukherjee and D. F. Agterberg}
\address{Department of Physics, University of Wisconsin-Milwaukee, Milwaukee, WI 53211}
\begin{abstract}
We examine weak-coupling theory for unconventional superconducting
states of cubic or tetrahedral symmetry for arbitrary order
parameters and Fermi surfaces and identify the stable states in
zero applied field. We further examine the possibility of having
multiple superconducting transitions arising from the weak
breaking of a higher symmetry group to cubic or tetrahedral
symmetry. Specifically, we consider two higher symmetry groups.
The first is a weak crystal field theory in which the spin-singlet
Cooper pairs have an approximate spherical symmetry. The second is
a weak spin orbit coupling theory for which spin-triplet Cooper
pairs have a cubic orbital symmetry and an approximate spherical
spin rotational symmetry. In hexagonal UPt$_3$, these theories
easily give rise to multiple transitions. However, we find that
for cubic materials, there is only one case in which two
superconducting transitions occur within weak coupling theory.
This sequence of transitions does not agree with the observed
properties of PrOs$_4$Sb$_{12}$. Consequently, we find that to
explain two transitions in PrOs$_4$Sb$_{12}$ using approximate
higher symmetry groups requires a strong coupling theory. In view
of this, we finally consider a weak coupling theory for which two
singlet representations have accidentally nearly degenerate
transition temperatures (not due to any approximate symmetries).
We provide an example of such a theory that agrees with the
observed properties of PrOs$_4$Sb$_{12}$.
\end{abstract}
\maketitle
\section{ INTRODUCTION}
It has been observed that a wide variety of heavy fermion
superconductors appear to undergo multiple phase transitions within
the superconducting state. These materials include
UPt$_3$\cite{upt1}, U$_{1-x}$Th$_x$Be$_{13}$\cite{uth1,uth2} and
PrOs$_4$Sb$_{12}$\cite{pro1,pro2}. The fact that multiple
transitions are being observed in such a significant fraction of the
total number of heavy fermion superconductors discovered is
presumably providing a valuable insight into the nature of the
superconducting state. Among the heavy fermion superconductors
showing multiple phase transitions, UPt$_3$ has been most
extensively studied. It has a hexagonal point group symmetry and
shows two transitions with a  small (about 10\%) splitting of the
superconducting transition temperature.
 Three main approaches have been used to theoretically model the
phase diagram of UPt$_3$ and it would be of interest to see if any
of these approaches apply to the other materials. The first approach
uses a weak symmetry breaking field (SBF) to break the hexagonal
symmetry. This field lifts the degeneracy of  a multi-component
order parameter leading to two transitions \cite{hess}. The origin
of this symmetry breaking field has been questioned\cite{luss} and
other proposals have emerged. Zhitomirsky and Ueda have postulated a
weak crystal field model in which spin singlet Cooper pairs
experience an approximate spherical symmetry which is weakly broken
to hexagonal symmetry \cite{zhit}. Along similar lines, Machida and
co-workers have postulated a weak spin orbit coupling for
spin-triplet Cooper pairs\cite{mach} . The third approach uses a
phenomenological model which considers two different irreducible
representations of the hexagonal point group that accidentally have
nearly the same $T_c$ \cite{chen,mineev} . In all these cases, it is
possible to develop microscopic theories based on weak coupling
theory that give rise to the two transitions and such theories have
been useful in developing a qualitative understanding of this
superconductor \cite{joynt}.

Among the materials with cubic or tetrahedral point group symmetry:
U$_{1-x}$Th$_x$Be$_{13}$ and PrOs$_4$Sb$_{12}$ have shown multiple
superconducting transitions. Both these materials have been the
subject of phenomenological studies. U$_{1-x}$Th$_x$Be$_{13}$ has
been studied using a phenomenological approach by Sigrist and Rice
\cite{uth3} while PrOs$_4$Sb$_{12}$ has been examined
phenomenologically by Goryo \cite{goryo}, Ichioka et.
al.\cite{ichi},  Matsunaga et. al.\cite{matsu}, and more recently by
Sergienko and Curnoe\cite{serg}. There have been a variety of
microscopic proposals for the superconducting states in
PrOs$_4$Sb$_{12}$ \cite{miyaki,maki,matsu1}.

 It is interesting to note that none of these microscopic
theory attempts to provide an origin for the two transitions for
cubic or tetrahedral materials. Existing microscopic theories
predict the properties of the A phase but are forced to make ad-hoc
assumptions about the origin of the transition in the B phase. This
is in sharp contrast to the case of UPt$_3$ where microscopic
justification for the second transition exist based on weak coupling
theories \cite{joynt}.\\

 In the hope of identifying a common origin to multiple transitions
 for heavy fermion superconductors, we apply
the conceptual frameworks developed for hexagonal UPt$_3$ to cubic
and tetrahedral superconductors. One goal is to examine under what
circumstances weak coupling BCS provides a microscopic description
of the two transitions.  We expect that BCS theory will capture a
large contribution to the condensation energy of these heavy fermion
superconductors and therefore provides a reasonable basis for such
studies. This is partially justified by the agreement between the
observed size of the specific heat jumps at $T_c$ and that predicted
by weak coupling theories. In neither U$_{1-x}$Th$_x$Be$_{13}$ nor
PrOs$_4$Sb$_{12}$ has any weak symmetry breaking fields been
identified. Therefore, we begin by exploring possible higher
symmetry groups that are weakly broken. We consider two higher
symmetries groups. The first is motivated by weak crystal field
theory of Zhitomirsky and Ueda for UPt$_3$\cite{zhit}. In this
theory, the spin-singlet superconducting state has an approximate spherical
symmetry $SO(3)$ which is weakly broken to cubic or tetrahedral
symmetry. We next consider the possibility of weak spin-orbit
coupling so that the spin-triplet superconductor has an approximate $O\times
SO(3)$ symmetry. We find that weak coupling theory for only the latter of these two
higher symmetries allow for one possible scenario for two
transitions. The sequence of transitions found this way does not
agree with experimental data for PrOs$_4$Sb$_{12}$. We further show
that the weak crystal field theory does allow for two transitions
when strong coupling corrections are included. However, the
condensation energy associated with the strong coupling corrections
must be comparable to that of weak coupling theory to account for
experimental results on the specific heat. These results indicate
 that a weak coupling theory does not provide an adequate description of the superconducting
 state within the context of weakly broken higher symmetry groups.

Given the failure of weak coupling theory for PrOs$_4$Sb$_{12}$ in
the above context, we finally ask if it is possible for weak
coupling theory to provide a description of the two transitions that
agrees with the experimental results. We find that it
is possible for such an approach to succeed and give one example of
when it does. However, without the constraints imposed by higher
symmetries, a general analysis of resulting weak coupling theories
is not possible and requires a detailed knowledge of the quasi
particles within the material.

The paper is organized as follows. We initially provide an overview
of experimental results for PrOs$_4$Sb$_{12}$ and
U$_{1-x}$Th$_x$Be$_{13}$. Then we provide an analysis that results
in all the possible high temperature superconducting phases allowed
within weak coupling theory for cubic and tetrahedral
superconductors. Finally, we examine the origin of the second transition
within the frameworks discussed above.
\section{ EXPERIMENTAL PROPERTIES OF THE SUPERCONDUCTORS PrOs$_4$Sb$_{12}$ AND U$_{1-x}$Th$_x$Be$_{13}$}

The heavy fermion superconductor PrOs$_4$Sb$_{12}$ is the first
among the rare-earth filled skutterudite compounds showing a
superconducting behavior. It undergoes a normal to superconducting
transition with $T_{c1}=1.85$ K (the high temperature phase is known as the A phase) followed by another
transition at $T_{c2}=1.75$ K from the A phase to the B phase. This second transition shows up as a pronounced anomaly in the specific heat and magnetization
measurements\cite{pro3,pro4}. Note that the
specific heat measurements by Measson {\it et al.} have raised questions
about the intrinsic nature of the
double superconducting transition\cite{pro5}. Various experiments have reported a
similar field dependance of the H$_{c2}$ curve for both the
transition temperatures\cite{pro4}. Though the symmetry and type of
nodes of the gap structure in the A phase remains inconclusive,
experiments suggest the presence of two point nodes in the B phase.
This has been observed by the power law temperature dependance in
specific heat\cite{pro2}, thermal conductivity \cite{pro6} and
penetration depth measurements\cite{pro7}. A possible third
phase transition($T_{c3})$ has been observed as an enhancement of
the lower critical field $H_{c1}(T)$ below T$\approx0.6$ K\cite{pro9}.
Interestingly this anomaly has not been detected by specific heat
measurements at low temperatures. Experiments have also observed
local broken time reversal symmetry in the superconducting phase through
$\mu$SR measurements\cite{pro8}.\\

One experimental result on PrOs$_4$Sb$_{12}$ which has not received
much theoretical attention yet provides a strong constraint is the
low field measurement of the vortex lattice geometry. The key result
is that the vortex lattice is not hexagonal near H$_{c1}$ for the
field along the $c$ axis\cite{profll}. As the authors of
\cite{profll} have pointed out, this is a strong constraint because
this lattice structure results from a London free energy whose form
depends upon the symmetry of the superconduting state. The London
free energy can be expanded in powers of the reciprocal lattice
vector $\vq$ of the vortex lattice. The form of the free energy is
\cite{kogan}
\begin{equation}
F={h^2\over8\pi}\sum_\mathbf{q}[1+\lambda^2\sum_{i,j}(1+m_{ii}q_j^2-m_{ij}q_iq_j)]
\end{equation}
here $\lambda$ is the penetration depth and $m$ is the normalized
London effective mass tensor whose components in weak coupling theory for a singlet superconductor is given by
$m^{-1}_{ij}\propto <v_{fi}v_{fj}|\Delta(\vk)|^2>_{FS}$ with $v_{fi}$ being
the $i^{th}$ component of Fermi velocity and $|\Delta(\vk)|$ representing
the gap magnitude. We are justified in keeping $\vq$ terms up to
second order near H$_{c1}$ because in this region $\vq$ has a small
magnitude being approximately given by
$q\approx\sqrt{B\over\phi_0}$.
If we consider that the gap to be invariant under a three fold
rotation we find that the components of the effective mass tensor
are given by
$$m_{xx}=m_{yy}=m_{zz},m_{xy}=m_{yz}=m_{xz}=0$$
This situation would result in a hexagonal
 vortex lattice near H$_{c1}$. No other symmetry element of the tetrahedral point group implies a hexagonal vortex lattice. Since the observed vortex lattice is not hexagonal near $H_{c1}$, we conclude
 that superconducting state  does not contain a three fold symmetry
 axis.

The alloy U$_{1-x}$Th$_x$Be$_{13}$ which has cubic point group
symmetry shows two second order superconducting transition for
concentration of thorium exceeding $x=0.018$ in specific heat
measurements \cite{ott}. A pronounced peak has been observed in the
ultrasonic attenuation for longitudinal sound propagated along a
[100] direction at $T=T_{c2}$\cite{bat}. Measurements by Bishop {\it et
al.} found a similar behavior for longitudinal sound propagated
along the [111] direction\cite{bishop}. Significant anomalies have
been observed in $H_{c1}(T)$ at $T=T_{c2}$\cite{rouch} and the zero
field $\mu$SR line width shows a marked increase as $T$ decreases
below $T_{c2}$ in samples with $x\approx0.033$\cite{heff}. Finally
Lambert {\it et al.} found differences in the pressure dependance of $T_c$
for $x<0.018$ and $x>0.018$\cite{lamb}.

\section{Theory of the High Temperature Superconducting Phase}
 For all the theories we consider, the initial transition
into the superconducting state is characterized by either cubic or
tetrahedral symmetry. Therefore, this transition is described by a
single order parameter symmetry. Here, we provide a general analysis
of  the possible superconducting states that weak-coupling theory allows
for this phase. A similar analysis of this problem has appeared
recently by Kuznetsova and Barzykin\cite{bar}. However, we find that
weak coupling theory provides even stronger constraints than found
in this work. As a consequence, some of the phases found by these
authors are ruled out. \\
 The cubic group O$_h$ has ten irreducible representations (irreps) that are listed in
 Table~\ref{table1}. The tetrahedral group has one three dimensional irrep and three one dimensional
 irreps, two of which are complex conjugate and combine to form a single irrep when time reversal symmetry is present.  In this Section we study the
superconducting phases of materials with cubic and tetrahedral point
group symmetry whose basis functions transform as a multidimensional
representation.
\begin{center}
\begin{table}
\begin{tabular}{|c|c|c|c|}\hline
Representation   &Representative basis function($f$) &Representation
& Representative basis function($f$)\\ \hline $A_{1g}$ & $
k_x^2+k_y^2+k_z^2$ & $A_{1u}$ & $ \mhx k_x+\mhy k_y+\mhz
k_z$\\\hline $A_{2g}$ & $ (k_x^2-k_y^2)(k_y^2-k_z^2)(k_z^2-k_x^2)$ &
$A_{2u}$ & $ \mhx k_x(k_z^2-k_y^2)+\mhy k_y(k_x^2-k_z^2) +\mhz
k_z(k_y^2-k_x^2)$\\\hline $E_g$ & $ 2k_z^2-k_x^2-k_y^2,k_x^2-k_y^2$
& $ E_u$ & $2\mhz k_z-\mhx k_x-\mhy k_y,\mhx k_x-\mhy k_y$\\\hline
$T_{1g}$ & $
k_yk_z(k_y^2-k_z^2),k_zk_x(k_z^2-k_x^2),k_xk_y(k_x^2-k_y^2)$ &
$T_{1u}$ & $\mhy k_z-\mhz k_y,\mhz k_x-\mhx k_z,\mhx k_y-\mhy
k_x$\\\hline $T_{2g}$ & $ k_yk_z,k_xk_z,k_xk_y$ & $T_{2u}$ & $\mhy
k_z+\mhz k_y,\mhz k_x+\mhx k_z,\mhx k_y+\mhy k_x$\\\hline
\end{tabular}
\caption[Table 1.]{The irrep and corresponding representative basis
functions for even and odd parity states of cubic symmetry}
\label{table1}
\end{table}
\end{center}
The superconducting gap is given by $\Delta(\vk) = [ \psi(\vk)
\sigma_0 + \vd (\vk) \cdot \vsig ] i \sigma_y $ where $
\psi(\vk)=\psi(-\vk)$ is the even-parity spin-singlet component, $
\vd(\vk)=-\vd(-\vk)$ is the odd-parity spin-triplet component, and
$\vsig=(\sigma_x,\sigma_y,\sigma_z)$ represent the Pauli spin
matrices. Among the various irreps of the point group symmetry there
is one, say $\Gamma$, which gives the highest transition
temperature. The superconducting state can be written as a linear
combination of the basis functions of this representation
$\Delta(\Gamma,m;\vk)$
$$\Delta(\vk)=\sum_{m}\eta(\Gamma,m)\Delta(\Gamma,m;\vk)$$
Here $\eta(\Gamma,m)$ are in general complex and act as the order
parameter. For a single representation
the fourth order free energy for the superconducting state can be written as an expansion in $\eta(\Gamma,m)$\\
$$F_\Gamma(T,\eta)=F_0(T)+\alpha\sum_{m}|\eta(\Gamma,m)|^2+f_\Gamma[\eta(\Gamma,m)^4]$$
with $\alpha=\alpha_0[{T/ T_c(\Gamma)}-1]$, and the fourth order
energy contains all terms which are invariant under $G\times R\times
U(1)$. The normal state of the system is represented by the free
energy $F_0(T)$. Here $G$ is the crystal point group symmetry, $R$
and $U(1)$ are the time reversal and gauge symmetry groups
respectively. The fourth order terms in the Ginzburg Landau theory
are characterized by several coefficients $\beta_{\dot{\iota}}$
which are arbitrary in a general theory but are constrained in the weak coupling limit.\\
 For the spin triplet systems the fourth order free energy within weak-coupling theory can be evaluated
 from the average
 $$\left<|\vd(\vk)|^4+|\vq(\vk)|^2\right>$$
with $\vq=i\vd(\vk)\times\vd(\vk)^\ast$. The {\vq} vector is zero
for unitary states and takes finite values for non unitary states.
Since $|\vq|^2$ gives a positive fourth order contribution to the
free energy for non unitary states it is unusual for these states to
be preferred stable states within a weak coupling theory. For the spin-triplet representations, sixth order terms in the free energy are required to remove a residual degeneracy and completely specify the solution. The sixth order free energy is given by
$${1\over3}\left<|\vd(\vk)|^6+3|\vq(\vk)|^2|\vd(\vk)|^2\right>.$$
For spin singlet systems the fourth order free energy will be given
by
$$\left<|\psi(\vk)|^4\right>.$$
We now analyze the possible
superconducting ground states for cubic symmetry within the weak coupling limit.\\
\subsection{Two Dimensional Representation}
 The cubic group contains one two dimensional representation for
 even parity states with the gap function
$$\psi(\vk)=\eta_1f_{1E_g}(\vk)+\eta_2f_{2E_g}(\vk)$$
where  $f_{1E_g}(\vk)$ and $f_{2E_g}(\vk)$ form a basis for the
$E_g$ irrep. For the odd parity states the representative basis
functions $\vf_{1E_u}$ and $\vf_{2E_u}$ are given in Table~\ref{table1}
and the gap function is given by
$$\vd(\vk)=\eta_1\vf_{1E_u}(\vk)+\eta_2\vf_{2E_u}(\vk)$$
The general free energy can be expressed as an expansion in the
order parameter $\eta_1$ and $\eta_2$ as
\begin{equation}
 F = \alpha_0\left[1-{T\over T_c}\right](|\eta_1|^2+|\eta_2|^2)+{7\zeta(3)\over16(\pi T_c)^2}N_0
 \Delta^4[\beta_1(|\eta_1|^2+|\eta_2|^2)^2
 +\beta_2(\eta_1^*\eta_2-\eta_2^*\eta_1)^2]
 \end{equation}
 Where $\alpha_0=1$ in the weak coupling limit. The weak coupling values of the fourth order coefficient for the
 even parity states are
 $$\beta_1=3\beta_2=<f_{1E_g}^4(\vk)>$$
 with the bracket meaning the average over the Fermi surface.
 All averages in the current and future discussion are in a
  normalized form such that
 $<f^2(\vk)>=1$ where $f(\vk)$ represents the basis function.
 If we minimize this free energy we find that since $\beta_1>0$, the
 phase $\omega^2(1,i)$ will minimize the free energy. This is true for arbitrary Fermi surfaces and gap basis functions. This phase belongs to the
 superconducting class $O(D_2)$ \cite{gor1} and has point nodes along the cube
 diagonals. The gauge factor $\omega^2$
 has been multiplied to keep the notation consistent with Ref. 36.

 For the odd parity irreps the weak coupling values for the
 fourth order coefficients are
  $$\beta_1=3<f_x^2(\vk)f_y^2(\vk)>(x+1) , \beta_2=<f_x^2(\vk)f_y^2(\vk)>(x-7)$$
  where $x=<f_x^4(\vk)>/<f_x^2(\vk)f_y^2(\vk)>$,
 and $[f_x(\vk),f_y(\vk),f_z(\vk)]$ form an arbitrary basis of $T_{1u}$ symmetry and have the same rotation properties
 as the vector $\vk$.
 If we minimize this free energy within the weak coupling limit,
 the non magnetic phase $\vupsilon=(1,0)$ belonging to the superconducting class
 $D_4\otimes R$ is found to be stable for $x<7$ (note that a residual continuous degeneracy remains for which any real combination of the two components minimizes the fourth order free energy, this is lifted by sixth-order terms). For $x>7$ the
 magnetic phase $\vupsilon=\omega^2(1,i)$ is stable. Again the gauge factor $\omega^2$ has been multiplied to keep the notation consistent with Ref.36.
  This phase belongs to the $O(D_2)$ superconducting
 class and contains point nodes along the cube diagonals.
  It is interesting to find a non unitary phase that is
 stabilized within weak coupling theory in zero applied field. The second non-magnetic phase $\vupsilon=(0,1)$ appears to be prohibited in a weak coupling theory
 by the sixth order term in the free energy
 $$-\gamma_3|\eta_1|^2|3\eta_2^2-\eta_1^2|^2$$
 where $\gamma_3$ is given by the weak coupling value
 $$\gamma_3={1\over54}<f_x^6(\vk)+2f_x^2(\vk)f_y^2(\vk)f_z^2(\vk)-3f_x^4(\vk)f_y^2(\vk)>$$
 we numerically find $\gamma_3>0$ for a variety of basis functions and Fermi surface structures, but we could not prove this  analytically.

 Tetrahedral symmetry does not change the structure of the  $\omega^2(1,i)$, but does introduce an additional sixth order term in
 the free energy that modifies the $\vupsilon=(1,0)$ phase. This term is
$${\sqrt{3}\over72}<f_x^2(\vk)f_y^2(\vk)(f_x^2(\vk)-f_y^2(\vk))>
[3(\eta_1\eta_2^*+\eta_2\eta_1^*)(|\eta_1|^4+|\eta_2|^4-3|\eta_1|^2|\eta_1|^2)-
(\eta_1^3{\eta_2^*}^3+\eta_1^3{\eta_2^*}^3)>$$
 as a result the stable A phase ground state is given by $(\phi_1,\phi_2)$, where both $\phi_1$ and $\phi_2$ are real
 (note that there is no continuous degeneracy in this phase) \cite{serg}.

 \subsection{Three Dimensional Representation}
 The Free energy for the three dimensional
 representation can be written as
 \begin{eqnarray}
 F&=& \alpha_0\left[1-{T\over T_c}\right](|p_1|^2+|p_2|^2+|p_3|^2)
 +{7\zeta(3)\over16(\pi T_c)^2}N_0 \Delta^4[\beta_1(|p_1|^2+|p_2|^2+|p_3|^2)^2\nonumber\\
 & & +\beta_2(p_1^2+p_2^2+p_3^2)({p_1^*}^2+{p_2^*}^2+{p_3^*}^2)+\beta_3(|p_1|^4+|p_2|^4+|p_3|^4)]\nonumber\\
  \end{eqnarray}
  The weak coupling values of the
  coefficients are
$$\beta_1=2\beta_2=2<f_{1T_{2g}}^2(\vk)f_{2T_{2g}}^2(\vk)>,\beta_3=<f_{1T_{2g}}^4(\vk)>-3<f_{1T_{2g}}^2(\vk)f_{2T_{2g}}^2(\vk)>$$
  The functions $[f_{1T_{2g}}(\vk),f_{2T_{2g}}(\vk),f_{3T_{2g}}(\vk)]$ form a basis for the $T_{2g}$ irrep.
  If we define the parameter space in terms of one free parameter
  $$\tilde{x}={<f_{1T_{2g}}^4(\vk)>\over<f_{1T_{2g}}^2(\vk)f_{2T_{2g}}^2(\vk)>}$$
  we find that for $\tilde{x}<3$, the state $(1,i,0)$ is the ground state.
  This state belongs to the class $D_4(E)$ and contains line nodes in the $z=0$ plane. For $\tilde{x}>3$,
  the phase $(1,\omega,\omega^2)$ is stable high temperature phase. This state
  belongs to the class $D_3(E)$ and contains point nodes. The boundary $\tilde{x}=3$ corresponds to a spherical
  Fermi surface. The weak coupling values provide a tighter restriction on the allowed phases than found by
 Kuznetsova {\it et. al.} and result in only magnetic states
being the stable weak coupling phases. In particular, it should be noted that $\tilde{x}>1$ for any choice of basis functions, and this constraint rules out any non-magnetic phases.

For the odd parity states the weak coupling values are given by
$$\beta_1=x+3,\beta_2=-{(x+1)\over2},\beta_3={x-3\over2}$$
where again $x=<f_x^4(\vk)>/<f_x^2(\vk)f_y^2(\vk)>$.\\
 We find that the state $(1,1,1)$ is stable
 for $x>3$ whereas $(1,0,0)$ is stable for $x<3$ for both $T_{1u}$ and $T_{2u}$ irreps. The boundary $x=3$ corresponds to the
 spherical Fermi surface.

 All the possible solutions for the high temperature phase within
 the weak coupling limit are listed in table~\ref{table2}. For the
 tetrahedral symmetry the analysis will be similar to the cubic at the fourth order as
 tetrahedral and cubic group have the same invariants. The
 difference in the tetrahedral and cubic invariants at the sixth
 order results in the state $(|\eta_1|,i|\eta_2|,0)$ belonging to
 $D_2(E)$ symmetry group giving the ground state for the tetrahedral symmetry rather
 than the $(1,i,0)$ state belonging to the $D_4(E)$ symmetry group
 giving the ground state for cubic symmetry\cite{serg}.

 In view of the controversy between the extrinsic versus intrinsic
nature of the transition in PrO$_4$Sb$_{12}$, we briefly consider
that there is single superconducting transition and ask if any of
the above stable phases can explain the experimental properties
observed at low temperatures. In particular, the ground states
listed in table~\ref{table2} which may explain the observed
properties are the $(1,0)$ state of $E_u$, $(1,i,0)$ for $T_{2g}$,
and the $(1,0,0)$ state of $T_{1u}$ and $T_{2u}$. These states have
a gap structure with point nodes or are highly anisotropic. In
addition, each of these states break tetrahedral symmetry which
would result in a distorted vortex lattice structure. Note that the
observation of an increased muon relaxation rate in the
superconducting phases \cite{pro8} does not imply that time reversal
symmetry is globally broken, but only locally broken \cite{sig1}. We
take this to imply that the order parameter must be multi-component
but that ordered phase need not break time reversal symmetry
globally.

 \begin{center}
\begin{table}
\begin{tabular}{|c|c|c|c|}\hline
Representation   & State &Nodes   & Symmetry Class\\
\hline $E_g$ & $ {\omega^2\over\sqrt{2}}(1,i)$ & $P$ & $
O(D_2)$\\\hline
 $E_u$ & $(1,0)$ & $-$ & $D_4\otimes R$\\\hline
 $E_u$ & ${\omega^2\over\sqrt{2}}(1,i)$& $ P$ & $O(D_2)$\\\hline
$T_{2g}$ & ${1\over\sqrt{2}}(1,i,0)$ & $L$ & $D_4(E)$\\\hline
$T_{2g}$ &${1\over\sqrt{3}}(1,\omega,\omega^2)$ & $P$ &
$D_3(E)$\\\hline $T_{2u}$ & ${1\over\sqrt{3}}(1,1,1)$& $ P$ &
$D_3\otimes R$\\\hline $T_{2u}$ & $(1,0,0)$& $ P$ & $D_4(D_2)\otimes
R$\\\hline $T_{1u}$ & ${1\over\sqrt{3}}(1,1,1)$& $ P$ &
$D_3(C_3)\otimes R$\\\hline $T_{1u}$ & $(1,0,0)$& $ P$ &
$D_4(C_4)\otimes R$\\\hline
\end{tabular}
\caption[Table 2.]{Stable high temperature phases in the weak
coupling limit. The corresponding representative basis functions are
listed in table~\ref{table1}. Here $\omega=\exp[2\pi i/3]$}
\label{table2}
\end{table}
\end{center}
\section{Weakly broken $SO(3)$ theory for spin singlet states}

We now turn to the role that higher symmetries may play in giving
rise to two superconducting phase transitions. In this section, we
consider the weak crystal field theory for which there is an
approximate $SO(3)$ for spin-singlet superconductors. Such an
approach was proposed to explain the phase diagram of UPt$_3$ by
Zhitomirsky, and Ueda \cite{zhit}.

We find the possible superconducting transitions for a  state in
which the spin singlet Cooper pairs are in the $l=2$ channel. Due to the
effect of a weak crystal field, the five-fold degenerate $l=2$ irrep
of $SO(3)$ split into $E_g\oplus T_{2g}$ of the cubic group. The free energy for the $l=2$ irrep of $SO(3)$ has been found by Mermin and Stare \cite{mer}, incorporating the weak cubic field gives
\begin{equation}
f=\alpha_1(|\eta_1|^2+|\eta_2|^2)+\alpha_2(|p_1|^2+|p_2|^2+|p_3|^2)+
\beta_1|TrB^2|^2+\beta_2(TrB^*B)^2+\beta_3Tr(B^2{B^*}^2)
\end{equation}
  here $B$ is a $3\times3$ traceless
  symmetric complex
matrix given by the $l=2$ order parameter $\psi(\vk)=\sum_{\mu
\nu}B_{\mu \nu}k_\mu k_\nu$. Note that since the symmetry breaking
is weak, spherical symmetry is broken by the second order term only.
The weak coupling limit corresponds to the special case,\cite{mer}
 $\beta_2=2\beta_1$, $\beta_3=0$. The magnitude of $\beta_3$ is
only due to strong coupling effects and is of order $T_c/E_F$
. The general form of the gap is\\
 $$\psi(\vk)={\eta_1\over \sqrt{6}}(2k_z^2-k_x^2-k_y^2)+{\eta_2\over \sqrt2}(k_x^2-k_y^2)+\sqrt{2}p_1k_xk_y+\sqrt{2}p_2k_yk_z+\sqrt{2}p_3k_zk_x$$
 here $\mbox{\boldmath$\eta$}=(\eta_1,\eta_2)$ transform like the $E_g$ representation and $\vP=(p_1,p_2,p_3)$ transform like the $T_{2g}$
 representation of the cubic group. The components of matrix B can be written as
$$B_{xx}={\eta_1\over\sqrt{3}}+\eta_2,B_{yy}={\eta_1\over\sqrt{3}}-\eta_2,B_{zz}=-{2\eta_1\over\sqrt{3}},B_{xy}=p_1,B_{yz}=p_2,B_{zx}=p_3$$
 \begin{center}
\begin{table}
\begin{tabular}{|c|c|}\hline
$\Gamma$   & $F_4$\\ \hline $\mbox{\veta}=(1,0),\vP=0$ & $
\beta_1+\beta_2+\beta_3/2$\\\hline
$\veta={\omega^2\over\sqrt{2}}(1,i),\vP=0$ & $
\beta_2+\beta_3/3$\\\hline $\veta=0,\vP={1\over\sqrt{2}}(1,i,0)$ & $
\beta_2+\beta_3/4$\\\hline $\veta=0,\vP=(1,0,0)$ & $
\beta_1+\beta_2+\beta_3/2$\\\hline
$\veta=0,\vP={1\over\sqrt{3}}(1,\omega,\omega^2)$ & $
\beta_2+\beta_3/4$\\\hline $\veta=0,\vP={1\over\sqrt{3}}(1,1,1)$ & $
\beta_1+\beta_2+\beta_3/2$\\\hline
\end{tabular}
\caption[Table 3.]{The fourth order free energies ($F_4$) for
possible A phase representation ($\Gamma$) for spherical symmetry.
Here $\veta=(\eta_1,\eta_2)$, $\vP=(p_1,p_2,p_3)$ transform as
irreps of $E_g$ and $T_{2g}$ respectively of cubic symmetry. Here
$\omega=\exp[2\pi i/3]$} \label{table3}
\end{table}
\end{center}
We now look for the various transitions into the B phase which can
result in further stable superconducting transitions. The form of
this theory is simple enough to enable us to perform a general
analysis beyond the weak coupling limit. We will therefore look for
all possible transitions from all possible stable A phase solutions
listed in \cite{sig1}. From the values of the fourth order energies
in Table~\ref{table3} we find that for $\beta_2>0$, $\beta_3>0$ the
states $\vP=(1,i,0)$ and $\vP=(1,\omega,\omega^2)$ will tend to
stabilize deep in the B phase when the second order coefficients can
be ignored whereas for $\beta_2>0$, $\beta_3<0$ the state
$\veta=\omega^2(1,i)$ tends to stabilize. Within the weak coupling
theory these states do not give any stable second order transitions
into the B phase. If we include strong coupling effects
($\beta_3\neq0$) we find that there is only one second order
transition for $\beta_2>0,\beta_3<0$ from the state
$\vP={r\over\sqrt{2}}(1,i,0)$ in the A phase. This transition
corresponds to a B phase given by a linear combination of $\vP={|p|\over\sqrt{2}}(1,i,0)$
and $\veta=|\eta|e^{\dot{\imath}\pi\over2}(1,0) $ with a transition
temperature
$$T_{cB}=T_{cA}+3(1+{4\beta_2\over\beta_3})(T_{cA}-T_<)$$ where $T_<$ is
the transition temperature corresponding to pure
$\veta=(\eta_1,\eta_2)$ state. This state is highly anisotropic
and gives a distorted vortex lattice structure. It is therefore a
possible transition sequence for PrOs$_4$Sb$_{12}$. The specific
heat jump ratio between the transition at the B phase to the
transition at the A phase is
$${C_B\over C_A}={T_{cB}\over
T_{cA}}{\beta_3^2(\beta_2+\beta_3/4)\over144(1+\beta_2+\beta_3/2)(\beta_2+\beta_3/3)^2}$$
This transition gives a specific heat jump ratio of the order
$\beta_3^2$ which is negligible close to the weak coupling limit.
It is also interesting to note that this transition corresponds to
a change in penetration depth of order $\beta_3$ which can be a
significant change to observe in an experiment. A similar
explanation may hold for the $T_{c3}$ in PrOs$_4$Sb$_{12}$
observed in penetration depth measurements below T=0.6K but not
yet observed in specific
heat measurements.\\
\section{Weakly broken $O\times SO(3)$ theory for spin-triplet states}
We will now analyze the transition for the spin triplet states
within a weak coupling theory. We consider the effects of a weak
spin orbit coupling and also include the crystal field with cubic
symmetry and allow the spin channel to be isotropic. The irreps of
the symmetry group are given by the combined group $O_h\times
SO(3)\times R$. If we consider a weak spin orbit coupling in our
system, the basis functions split up into four different irreps. We
write the vector gap equation in terms of these irreps as
$\vd(\vk)=\Sigma_i\vd^{\Gamma_{i}}(\vk)$ where the components are
given by
\begin{eqnarray}
\vd^{A_{1u}}(\vk)&=&\lambda{1\over{\sqrt{3}}}[\mhx f_x(\vk)+\mhy f_y(\vk)+\mhz f_z(\vk)],\nonumber\\
\vd^{E_u}(\vk)&=&
\upsilon_1{1\over{\sqrt{2}}}[\mhx f_x(\vk)-\mhy f_y(\vk)]+\upsilon_2{1\over{\sqrt{6}}}[-2\mhz f_z(\vk)+\mhx f_x(\vk)+\mhy f_y(\vk)]\nonumber\\
\vd^{T_{1u}}(\vk)&=&p_1{1\over{\sqrt{2}}}[\mhy f_z(\vk)-\mhz
f_y(\vk)]+p_2{1\over{\sqrt{2}}}[\mhx f_y(\vk)-\mhy f_x(\vk)]
+p_3{1\over{\sqrt{2}}}[\mhz f_x(\vk)-\mhx f_z(\vk)]\nonumber\\
\vd^{T_{2u}}(\vk)&=& q_1{1\over{\sqrt{2}}}[\mhy f_z(\vk)+\mhz
f_y(\vk)]+q_2{1\over{\sqrt{2}}}[\mhx f_y(\vk)+\mhy
f_x(\vk)]+q_3{1\over{\sqrt{2}}}[\mhz f_x(\vk)+\mhx f_z(\vk)]
\nonumber\\
\end{eqnarray}
the components of the vector gap equation can be written in terms
of the basis
$\vd(\vk)=\sum_j\veta_jf_j(\vk)$, and we will use both bases for convenience.

For a weak spin orbit coupling, we get a free energy of the form
 \begin{eqnarray}
 F&=&\alpha_1|\lambda|^2+\alpha_2(|\upsilon_1|^2+|\upsilon_2|^2)+
 \alpha_3(|p_1|^2+|p_2|^2+|p_3|^2)+\alpha_4(|q_1|^2+|q_2|^2+|q_3|^2)
 +\beta_1(|\veta_x|^4+|\veta_y|^4+|\veta_z|^4)\nonumber\\& &+\beta_2(|\veta_x^2|^2+
 |\veta_y^2|^2+|\veta_z^2|^2)
 +\beta_3(|\veta_x.\veta_y|^2
 +|\veta_y.\veta_z|^2+
 |\veta_x.\veta_z|^2)+\beta_4[(\veta_x.\veta_y^*)^2
 +(\veta_x.\veta_z^*)^2+(\veta_y.\veta_z^*)^2
 +c.c]
 \nonumber\\& &+\beta_5(|\veta_x.\veta_y^*|^2+|\veta_y.\veta_z^*|^2
 +|\veta_x.\veta_z^*|^2)+\beta_6[(\veta_x)^2(\veta_y^*)^2+
 (\veta_x)^2(\veta_z^*)^2+(\veta_y)^2(\veta_z^*)^2+c.c]\nonumber\\& &+\beta_7
 (|\veta_x|^2|\veta_y|^2+
 |\veta_y|^2|\veta_z|^2+|\veta_z|^2|\veta_x|^2)
 \end{eqnarray}
  The weak coupling values of the normalized coefficients are$$\beta_1=-2\beta_2=<f_x^4(\vk)>,\beta_4={1\over2}\beta_5=-2\beta_6={1\over2}\beta_7=-{1\over2}\beta_3=<f_x^2(\vk)f_y^2(\vk)>$$

 \begin{center}
 \begin{table}
\begin{tabular}{|c|c|c|}\hline
$\Gamma$   & $F_4$  &W.C  \\
\hline $\lambda=1$  & $(\beta_1+\beta_2+\beta_7+2\beta_6)/3$
&$(x+2)/6$\\\hline $\vupsilon=(1,0),\vP=(1,0,0),\vQ=(1,0,0)$& $
(\beta_1+\beta_2+\beta_6)/2+\beta_7/4$
&$(x+1)/4$\\\hline$\vupsilon={\omega^2\over\sqrt{2}}(1,i)$ &
$(\beta_1+\beta_2-\beta_6+\beta_7)/3$ &$(5+x)/6$\\\hline
$\vP={1\over\sqrt{3}}(1,\omega,\omega^2)$ & $
(\beta_1+\beta_7)/3+(\beta_2+\beta_3+\beta_5)/12-\beta_4/12$
&$(15+7x)/24$\\\hline $\vQ={1\over\sqrt{3}}(1,\omega,\omega^2)$ &
$(\beta_1+\beta_7)/3+(\beta_2+\beta_3+\beta_5)/12-\beta_4/12$
&$(15+7x)/24$\\\hline $\vP={1\over\sqrt{2}}(0,1,i)$ & $
(3\beta_1+\beta_2-\beta_4-\beta_6)/8+(\beta_3+\beta_5+5\beta_7)/16$
&$(5x+9)/16$\\\hline $\vQ={1\over\sqrt{2}}(0,1,i)$ & $
(3\beta_1+\beta_2-\beta_4-\beta_6)/8+(\beta_3+\beta_5+5\beta_7)/16$
&$(5x+9)/16$\\\hline $\vP={1\over\sqrt{3}}(1,1,1)$ & $
(\beta_1+\beta_7)/3+(\beta_2+\beta_3+\beta_5)/12-\beta_4/12$
&$(x+3)/6$\\\hline $\vQ={1\over\sqrt{3}}(1,1,1)$ &
$(\beta_1+\beta_7)/3+(\beta_2+\beta_3+\beta_5)/12-\beta_4/12$
&$(x+3)/6$\\\hline
\end{tabular}
\caption[Table 4.]{The fourth order free energies($F_4$) for
possible A phase irreps($\Gamma$) and their corresponding weak
coupling values(W.C). The weak coupling values are in units of
$<f_x^2(\vk)f_y^2(\vk)>$. The value of $x$ is given by the ratio
${<f_x^4(\vk)>\over<f_x^2(\vk)f_y^2(\vk)>}$. Note that $x$=3
corresponds to spherical symmetry. $\omega=\exp[2\pi i/3]$.}
\label{table4}
\end{table}
\end{center}
It may be seen from Table~\ref{table3} that for $x>3$ the states
$\vP=(1,1,1)$ and $\vQ=(1,1,1)$ will minimize the fourth order terms
whereas for $x<3$, $\vupsilon=(1,0),\vP=(1,0,0),\vQ=(1,0,0)$ and
equivalent states minimize the fourth order free energy. At $x=3$ we
have a spherical fermi surface.

To understand if phase transitions are possible we compare the
states that minimize the second order term with those that minimize
the fourth order terms. If these are different, then a transition is
possible. However many of these transitions are first order. For
example the non unitary state $\vupsilon=\omega^2(1,i)$ which is
stable in the A
 phase for $x>7$ undergoes a first order transition to a mixed state
 with $\vupsilon=\omega^2(1,i)$ and $\vP=-i(1,\omega,\omega^2)$.\\
  We find that there is only one stable second order transition into the B
  phase.
  This instability is from $\vP=p(1,1,1)$ to the combination of
  $\vP=p(1,1,1)$ and $\vQ=q(1,1,1)$, where $p$ and $q$ are the order
  parameter values.
 Within the assumption of a small spin orbit coupling we write the general form of the gap
 for the states $\vP=(1,1,1)$
  and $\vQ=(1,1,1)$ as
  \begin{eqnarray}
   \Delta(\mathbf{k})&=&\cos\theta([\mhx f_y(\vk)+\mhy f_x(\vk)]+[\mhz f_x(\vk)+\mhx f_z(\vk)]+[\mhy
   f_z(\vk)+\mhz+
   f_y(\vk)])\nonumber\\
   & &\sin\theta([\mhx f_y(\vk)-\mhy f_x(\vk)]+[\mhz f_x(\vk)-\mhx
   f_z(\vk)]+[\mhy f_z(\vk)-\mhz f_y(\vk)])
  \end{eqnarray}
 Here $\theta$ acts as the
order parameter.
   If we assume that the $\vP=(1,1,1)$ state has
   the highest transition temperature,
   there is a transition from a $\theta=0$ state in A
   phase for which $\theta$ becomes non zero and grows towards a fully gapped system at
   $\theta=\pi/4$ (at which the gap has the form $\mhx f_y(\vk)+\mhy f_z(\vk)+\mhz f_x(\vk)$). This gives a
   second order transition temperature
    $$T_{cB}=T_{cA}-{1\over2}(x+3)(T_{cA}-T_<)$$
   and the ratio of jumps in the specific heat at the
   transition temperatures of the A and the B phases is
   $${c_B\over c_A}={T_{cB}\over T_{cA}(x+3)}$$\\
   It should  be pointed out that for tetrahedral group there will be
   an additional bilinear coupling term in the free energy
   $$\alpha_m(\vP^*\cdot\vQ+\vP\cdot\vQ^*)$$
   which would smear out the transition.\\

   Though we get two transitions in this case, the sequence does not
   satisfy the observed physical properties in the skutterudite
   PrOs$_4$Sb$_{12}$. Owing to the tendency of this system towards a
   fully gapped state such a transition would give a vortex lattice structure
   which becomes hexagonal at the lower temperatures which is contrary to the
   experimental observations where the distortions from a hexagonal structure
   increase at low temperatures\cite{profll}.  In addition this state does not satisfy
   the nodal structure of the gap in the B phase as observed in
    thermal conductivity and magnetization measurements.\\

 \section{Weak coupling accidental degeneracy theories}
 Here we consider the accidental degeneracy between the $T_{2g}$ and
 $E_g$ irreps as an example, since as explained at the end of this
 section there are many theories that have multiple transitions.

 In weak coupling theory we can write the fourth order free energy by evaluating the
 average$$<|\psi(\vk)|^4>=<|\eta_1f_{1E_g}(\vk)+\eta_2f_{2E_g}(\vk)+p_1f_{1T_{2g}}(\vk)+p_2f_{2T_{2g}}(\vk)+p_3f_{3T_{2g}}(\vk)|^4>$$
 In the above expression $\veta=(\eta_1,\eta_2)$ and
 $\vP=(p_1,p_2,p_3)$ are the order parameter values and the representative basis functions $f$ are assumed real.
  The free energy expression is
\begin{eqnarray}
 F&=&\alpha_1(|\eta_1|^2+|\eta_2|^2)+\alpha_2(|p_1|^2+|p_2|^2+|p_3|^2)
  +\beta_1[(|\eta_1|^2+|\eta_2|^2)^2+{1\over3}(\eta_1\eta_2^{*}-\eta_2\eta_1^{*})^2]\nonumber\\
 & & +\beta_2(|p_1|^4+|p_2|^4+|p_3|^4)\nonumber\\
 & &
 +\beta_3[4(|p_1|^2|p_2|^2+|p_2|^2|p_3|^2+|p_3|^4|p_1|^2)+p_1^2p_2^{*2}+p_2^2p_3^{*2}+p_3^2p_1^{*2}+c.c]\nonumber\\
 & &  +\beta_4/2[4(|\eta_1|^2+|\eta_2|^2)(|p_1|^2+|p_2|^2+|p_3|^2)+
 ((\eta_1^2+\eta_2^2)(p_1^{*2}+p_2^{*2}+p_3^{*2})+c.c)]\nonumber\\
  & &  -\beta_5/2[4(|\eta_1|^2-|\eta_2|^2)(2|p_3|^2-|p_1|^2-|p_2|^2)-4\sqrt{3}(\eta_1\eta_2^{*}+\eta_1^{*}\eta_2)(|p_1|^2
  -|p_2|^2)+\nonumber\\
  &
  &[(\eta_1^2-\eta_2^2)(2p_3^{*2}-p_2^{*2}-p_1^{*2})-2\sqrt{3}\eta_1\eta_2(p_1^{*2}-p_2^{*2})+c.c]]
 \end{eqnarray}
 Within a weak coupling theory the normalized coefficients are given by the following cubic averages
 $$\beta_1=<f_{1E_g}^4(\vk)>,\beta_2=<f_{1T_{2g}}^4(\vk)>,\beta_3=<f_{1T_{2g}}^2(\vk)f_{2T_{2g}}^2(\vk)>,\beta_4=<[f_{1E_g}^2(\vk)+f_{2E_g}^2(\vk)]f_{1T_{2g}}^2(\vk)>$$
 $$\beta_5=<[f_{1E_g}^2(\vk)-f_{2E_g}^2(\vk)]f_{1T_{2g}}^2(\vk)>$$
  The spherical fermi surface corresponds to the special case
 $$\beta_1=\beta_2=3\beta_3={3\over2}\beta_4,\beta_5=0$$
\begin{center}
\begin{table}
\begin{tabular}{|c|c|}\hline
$\Gamma$   & $F_4$ \\
\hline $\veta=(1,0),\vP=0$  & $ \beta_1$\\\hline
$\veta={\omega^2\over\sqrt{2}}(1,i),\vP=0$ & $ 2\beta_1/3$\\\hline
$\veta=0,\vP={1\over\sqrt{2}}(1,i,0)$ & $
(\beta_2+\beta_3)/2$\\\hline $\veta=0,\vP=(1,0,0)$ & $
\beta_2$\\\hline $\veta=0,\vP={1\over\sqrt{3}}(1,\omega,\omega^2)$ &
$ \beta_2/3+\beta_3$\\\hline
$\veta={1\over\sqrt{2}}(0,1),\vP={1\over\sqrt{2}}(0,0,i)$ & $
1/4(\beta_1+\beta_2+2\beta_4-4\beta_5)$\\\hline
$\veta=0,\vP={1\over\sqrt{3}}(1,1,1)$ & $
\beta_2/3+2\beta_3$\\\hline
\end{tabular}
\caption[Table 5.]{The fourth order free energies($F_4$) for irreps
of stable states($\Gamma$) in the A phase } \label{table5}
\end{table}
\end{center}

From Table~\ref{table5} we find that for $\beta_5<0$ the lowest 3
fourth order energies in increasing order correspond to states
$\vP=(1,i,0),\vP=(1,\omega,\omega^2),\veta=(1,i)$. We do not find
any second order weak coupling transitions within this range. For
$\beta_5>0$ the lowest 3 fourth order energies in increasing order
correspond to states
$\veta=(1,i),\vP=(1,\omega,\omega^2),\vP=(1,i,0)$. In this case we
find only one stable second order transition from the state
$\vP=|r|(1,i,0)$ in the A phase to a B phase where it mixes to the
state $\veta=e^{i\pi/4}(|\eta_1|,|\eta_2|e^ {i\pi/2})$ with a
transition temperature
  $$T_{cB}=T_{cA}-{1\over1-{{2\beta_4-\sqrt{7}\beta_5}\over\beta_2+\beta_3}}(T_{cA}-T_<)$$
   The specific heat jump ratio at the transition temperatures is
   given by
   $${c_B\over c_A}={T_{cB}\over
T_{cA}}\left({7(\beta_2+\beta_3-2\beta_4+\sqrt{7}\beta_5)^2\over12\beta_1(\beta_2+\beta_3)-7(2\beta_4-\sqrt{7}\beta_5)}\right)$$

   We find that a specific heat jump ratio that is comparable to
   observed value for the skutterudite PrOs$_4$Sb$_{12}$ which is about one\cite{pro3}can
   be obtained if the gap functions contain substantial cubic
   anisotropy. In addition this situation will result in a anisotropic state with two fold degeneracy.
   This sequence of phase transition would also result in a distorted
   vortex lattice structure owing to the large anisotropy of this state as shown in Figure 1.
  \begin{figure}
   \epsfxsize=3.0 in \center{\epsfbox{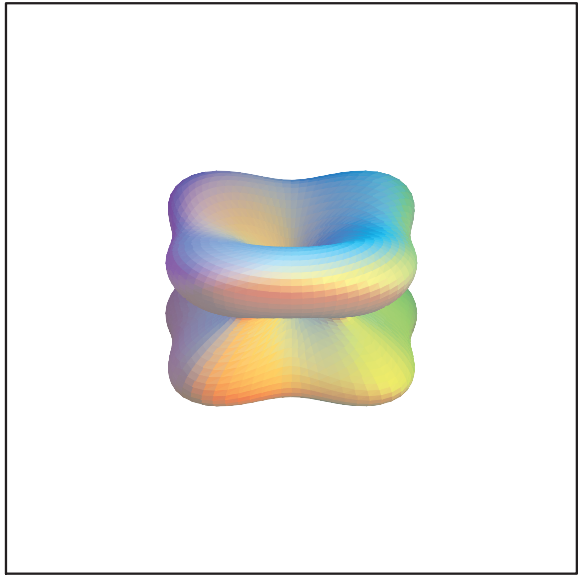}}\caption{(color online)Gap
   structure in the B phase for the
   $\mathbf{P}=|r|(1,i,0)+\mbox{\boldmath$\eta$}=e^{i\pi/4}(|\eta_1|,|\eta_2|e^
   {i\pi/2})$ phase.} \label{fig1}
    \end{figure}

     We find that in many
   cases, the A to B transition is first order. The reason for this
   is as follows:
   for two order parameters $\psi$ and $\eta$ the free energy takes
   the form
    $$F=\alpha_1|\psi|^2+\alpha_2|\eta|^2+\beta_1|\psi|^4+\beta_2|\eta|^4
     +\beta_{m1}|\psi|^2|\eta|^2+\beta_{m2}(\psi^2{\eta^*}^2+\eta^2{\psi^*}^2)$$
   Let $(\psi,\eta)=(|\psi|,|\eta|e^{i\phi})$ and minimize with
   respect to
   $\phi$. The the free energy becomes
   $$F=\alpha_1|\psi|^2+\alpha_2|\eta|^2+\beta_1|\psi|^4+\beta_2|\eta|^4+\beta_m|\eta|^2|\psi|^2$$
   where
   $\beta_m=\beta_{m1}-2\beta_{m2}\approx<f_\psi^2(\vk)f_\eta^2(\vk)>$.
   Here we have assumed the basis functions $f_\psi(\vk)$ and
   $f_\eta(\vk)$ to be real.
   If there is a second transition then it is second order transition when
   $\beta_m^2<4\beta_1\beta_2$, otherwise it is first order.
   In our calculations $\beta_m$ is too large, which leads to the
   first order transitions. This is a consequence of the functions
   $f_\psi(\vk)$ and $f_\eta(\vk)$ that have been chosen. However we
   can get a second order transition by considering a two band
   theory for which $f_\psi(\vk)$ is large and $f_\eta(\vk)$ is small in one
   band while $f_\eta(\vk)$ is large and $f_\psi(\vk)$ is small on
   the other. Then the condition $\beta_m^2<4\beta_1\beta_2$ will be
   easily satisfied and a second order A$\rightarrow$B phase
   transition can exist for almost any two different order parameter irreps.

\section{CONCLUSIONS}

 We have considered microscopic theories of unconventional superconductivity in cubic and tetrahedral superconductors.
 We have identified the stable weak-coupling unconventional superconducting states that belong to a single irreducible
 representation and have highlighted which of these can describe the low temperature properties of PrOs$_4$Sb$_{12}$.
 We have further examined theories for two intrinsic superconducting transitions in  PrOs$_4$Sb$_{12}$.
 We have found that a theory for which the two transitions are due to a weakly broken $SO(3)$ symmetry for spin singlet
 Cooper pairs cannot give rise to two transitions in the weak-coupling limit. However, it is possible for such a theory
 to produce two transitions that agree with the experimental properties of PrOs$_4$Sb$_{12}$ only when extended to the strong coupling limit.
 We further find that for spin-triplet Cooper pairs, weak spin-orbit coupling in the weak-coupling limit does not give rise
 to two superconducting transitions that agree with the experimental properties of PrOs$_4$Sb$_{12}$.
 Finally, we consider an example of a weak coupling theory that does not assume an approximate higher symmetry, but is
 based on the accidental closeness of the transition temperatures for two different
 representations. This example is able to describe the observed properties of PrOs$_4$Sb$_{12}$.

This work  was supported by the National Science Foundation grant No.
DMR-0381665.

\end{document}